\documentclass[twocolumn]{autart}
                \newcommand{\comments}[1]{}
                                                        \usepackage{natbib}

\usepackage{amsfonts,amssymb}
\usepackage{amsmath,amsfonts,amssymb,color}
\usepackage{psfrag}
\usepackage{epstopdf}

\newtheorem{theorem}{Theorem}
\newtheorem{proposition}{Proposition}
\newtheorem{remark}{Remark}

\newcommand{\rref}[1]{(\ref{#1})}

\usepackage{graphicx,color}

\usepackage{fancybox,latexsym,setspace,amssymb,amsfonts}
\usepackage{mathrsfs} 

\definecolor{pink}{rgb}{1,0.5,0.5} % color values Red, Green, Blue

\newcommand{\mm}[1]{}
\belowdisplayskip \abovedisplayskip%
\begin{document}
\begin{frontmatter}
\title{ Dwell-Time Based Stability Analysis  and $\mathcal{L}_2$ Control of LPV Systems with Piecewise
Constant Parameters and Delay\vspace{-.8em}}
%\thanks[footnoteinfo]{Corresponding author: F. Mazenc.\mm{ A preliminary version  has been submitted to the 2019 IEEE Conference on Decision and Control; see   Section \ref{sec:introduction} for   the differences between the conference version and this paper.} Malisoff was supported by US National Science Foundation Grant 1711299.}

%\author{ \parbox{3 in}{\centering Huibert Kwakernaak*
%         \thanks{*Use the $\backslash$thanks command to put information here}\\
%         Faculty of Electrical Engineering, Mathematics and Computer Science\\
%         University of Twente\\
%         7500 AE Enschede, The Netherlands\\
%         {\tt\small h.kwakernaak@autsubmit.com}}
%         \hspace*{ 0.5 in}
%         \parbox{3 in}{ \centering Pradeep Misra**
%         \thanks{**The footnote marks may be inserted manually}\\
%        Department of Electrical Engineering \\
%         Wright State University\\
%         Dayton, OH 45435, USA\\
%         {\tt\small pmisra@cs.wright.edu}}
%}

\author{Muhammad Zakwan}\ead{zakwan@ee.bilkent.edu.tr}\; \;
\author{Saeed Ahmed}\ead{ahmed@ee.bilkent.edu.tr}\; \;

\address{Department of Electrical and Electronics Engineering,
Bilkent University, Ankara 06800, Turkey}


\vspace{-.6em}
\begin{abstract}
Dwell-time based stability conditions  for a class of LPV systems with piecewise constant parameters under time-varying delay are derived using clock-dependent Lyapunov-Krasovskii functional. Sufficient synthesis conditions for clock-dependent gain-scheduled state-feedback controllers ensuring $\mathcal{L}_2$-performance are also provided.  Several numerical and practical examples, to illustrate the efficacy of the results, are given. 
\vspace{-1em}

\end{abstract}

\end{frontmatter}

%%%%%%%%%%%%%%%%%%%%%%%%%%%%%%%%%%%%%%%%%%%%%%%%%%%%%%%%%%%%%%%%%%%%%%%%%%%%%%%%

\spacing{1}
\section{Introduction}

The framework of LPV systems has proven to be a systematic way to model nonlinear real-world phenomena and synthesize gain-scheduled controllers for nonlinear systems; see \cite{BriatBook}, \cite{mohammadpour2012control}, and \cite{Toth}. The applications of LPV systems include the automotive industry (\cite{sename2013robust}), turbofan engines (\cite{gilbert2010polynomial}), robotics (\cite{kajiwara1999lpv}), and aerospace systems (\cite{shin2000h}). Apart from nonlinearity, real-world applications are often affected by time delays that can degrade the performance of the dynamical systems, or in the worst case, they can cause instability; see \cite{niculescu2001delay}. Time delays frequently appear in communication networks, mechanical systems, PVTOL aircrafts, robotized teleoperation, and many other domains; see \cite{chiasson2007applications} and \cite{ahmed2018dynamic}. Since time delays can also adversely affect the stability of the LPV systems (\cite{BriatBook}; \cite{zakwan2020distributedj}), it is quite natural to consider LPV systems with time delays.     

The point of view usually considered in LPV control is worst-case analysis, i.e., parameters are  assumed to behave
in an extreme way almost all the time by (i) either considering them to vary arbitrarily
fast/discontinuously, or (ii) by assuming that they have bounded derivatives. Both of them are quite extreme cases, and there is a room in the parameter space in-between parameters varying arbitrarily fast/discontinuously and parameters having bounded derivatives. To fill this gap, we consider the class of LPV systems with piecewise constant parameters as introduced in \cite{briat2015stability}. The rationale of LPV systems with piecewise constant parameters lies in reduced conservatism with improved performance. The main idea is to utilize the prior knowledge of the parameters' trajectory for stability analysis rather than performing worst-case analysis. LPV systems with piecewise constant parameters arise naturally in the context of sampled-data control of LPV
systems (\cite{joo2015lpv}) and control of buck converters with piecewise constant loads (\cite{tan2002output}). LPV systems with piecewise constant parameters can be considered as switched systems with an uncountable number of modes in a bounded compact set, \cite{zakwan2019poisson}.  LPV systems with piecewise constant parameters subject to spontaneous Poissonian jumps are also discussed in \cite{briat2018stability} and \cite{zakwan2019poisson}.  

The main aim of this paper is to study the dwell-time based stability properties and control of LPV systems with piecewise constant parameters under a time-varying delay. Stability analysis and control of LPV systems with piecewise constant parameters is also discussed in \cite{briat2015stability}. However, there are two main differences between our work and \cite{briat2015stability}.  First, no delay is present in \cite{briat2015stability}. Here we extend the results of \cite{briat2015stability} to the difficult case when there is a time-varying delay in the dynamics of LPV systems with piecewise constant parameters. Second, our work provides $\mathcal{L}_2$-performance for controller synthesis, which was not considered in \cite{briat2015stability}. 

At first glance, establishing quadratic stability for the class of LPV systems with piecewise constant parameters seems to be a natural choice, since the parameters belong to the class of arbitrarily fast varying parameters. However, by doing so, we will fail to capture the fact that the parameters are constant between the consecutive jumps, hence, leading to conservative results. To reduce this conservatism, we employ clock-dependent Lyapunov-Krasovskii functionals, introduced in \cite{briat2013convex}, for stability analysis.  These functionals inherit a clock that measures the time elapsed since the last jump in the parameters' trajectory yielding \emph{clock-dependent stability conditions}. These conditions result in infinite-dimensional semi-definite programs that are intractable. Several techniques such as gridding methods \citep[Appendix C]{BriatBook} and sum-of-squares (SOS) polynomials \citep{wu2005sos,scherer2006matrix} are available to approximate semi-infinite constraint LMI by a finite number of  LMIs. After obtaining the dwell-time based stability conditions, we also use them to derive synthesis conditions for 
clock-dependent gain-scheduled state-feedback controller ensuring $\mathcal{L}_2$-performance.  

\label{sec:introduction}

The paper unfolds as follows. In Section 2, we provide some preliminary results followed by dwell-time based stability conditions for LPV systems with piecewise constant parameters under a time-varying delay whereas synthesis conditions for clock-dependent gain-scheduled controllers with guaranteed $\mathcal{L}_2$ performance for these systems are provided in Section 3. Section 4 provides numerical and practical examples to illustrate our main results. Finally, some concluding remarks and future research directions are briefly discussed in Section 5.  

 We employ standard notation throughout the paper. The sets of positive integers and whole numbers are denoted by $\mathbb{N}$ and $\mathbb{N}_0 := \mathbb{N} \cup \{ 0 \}$, respectively. The identity and null matrices of dimension $n$ are denoted by $I_n$ and $\mathcal{O}_n$, respectively. We write $M\succ 0$ (resp. $M\preceq 0$) to indicate that $M$ is a symmetric positive definite (resp. negative semi-definite) matrix. The cone of  symmetric positive definite (resp. positive semi-definite) matrices is denoted by $\mathbb{S}^n_{\succ 0}\ (\text{resp.}\ \mathbb{S}^n_{\succeq 0})$.  For some square matrix $A$, $A + A^T$ will be denoted by $\textrm{Sym}[A]$. The Banach space of continuous functions from a set $X$ to a set $Y$ is denoted by $\mathscr{C}(X,Y)$.   The asterisk symbol $(*)$ denotes the complex conjugate transpose of a matrix and $x_t(\theta)$ is the shorthand notation for the translation operator acting on the trajectory such that $x_t(\theta) = x(t + \theta)$ for some non-zero interval $\theta \in [-h, 0]$. 

\section{Stability analysis of LPV systems with piecewise constant parameters and delay} 

\subsection{Preliminaries}
We consider in this paper LPV systems with piecewise constant parameters and time-varying delay that can be described as
\begin{equation*}
\label{sys_def}
\Sigma_{s}:\quad\left\lbrace \begin{array}{lll}
\dot{x}(t) &=& A(\rho)x(t) + A_d(\rho)x(t - d(t)) 
\\
&&+ B(\rho) u(t) + E(\rho) w(t) \\ 
z(t) &=& C(\rho)x(t) + C_d(\rho)x(t - d(t)) \\
&&+ D(\rho)u(t) + F(\rho)w(t) \\ 
x(\theta) &=& \phi(\theta), \ \forall \theta \in \left[-h, 0  \right]\; ,
\end{array}\right.
\end{equation*} 
where $x \in \mathbb{R}^n$ is the system state, $w \in \mathbb{R}^{m}$ is the exogenous input, $u \in \mathbb{R}^{q}$ is the control input,  $z \in \mathbb{R}^{r}$ is the controlled output,   and $\phi\in \mathscr{C}([-h \; 0], \mathbb{R}^n)$ is the functional initial condition. The time-varying delay $d(t)$  is assumed to belong to the set 
\[
\mathscr{D} :=  \lbrace d: \mathbb{R}_{\geq 0}\rightarrow [0,h],\ \dot{d} \leq \mu < 1 \rbrace\] with $h<+\infty$.
The parameter vector trajectory $ \rho : \mathbb{R}_{\geq 0} \rightarrow \mathcal{P} \subset \mathbb{R}^s$, $\mathcal{P}$ compact and connected, is assumed to be piecewise constant and measurable, and that the matrix-valued functions $A(\cdot)$, $A_d(\cdot)$, $B(\cdot)$, $C(\cdot)$, $C_d(\cdot)$, $D(\cdot)$, $E(\cdot)$, and $F(\cdot)$  are bounded and continuous on $\mathcal{P}$. We define the sequence $\{t_k \}_{k \in \mathbb{N}_0} , t_0 = 0$, of time instants where the parameters change values. We assume that there exists an $\epsilon > 0$ such that $T_k := t_{k+1} - t_k \geq \epsilon$ for all $k \in \mathbb{N}_0$, and $T_D \le T_k$ is refereed to as \emph{minimum dwell-time}.

We now provide the following integral inequality based on Jensen's inequality  to be used in the proof of our stability theorem to bound the derivative of the Lyapunov-Krasovskii functional. This inequality
plays an important role in the stability problem of time-delay systems, \cite{gu2003stability}.

\begin{proposition}[\cite{gu2003stability}] \label{prop:Jensen}  For any matrix valued function $R: \mathcal{P} \mapsto \mathbb{S}^n_{\succ 0}$, scalar $h>0$ such
that the integrations concerned are well defined, it holds that 
\begin{equation} \nonumber
\left(\displaystyle{\int^t_{t - h}} \dot{x}(s) ds\right)^T R \left(\displaystyle{\int^t_{t - h}}  \dot{x}(s) ds\right)\leq h \displaystyle{\int^t_{t - h}} \dot{x}(s)^T R \dot{x}(s) ds    \; .
\end{equation}
\end{proposition}
%{\em Proof:}\;
%\textcolor{blue}{Refer to \cite{gu2003stability}}. \hfill$\square$

\subsection{Dwell-time based stability results}

In this section, we derive dwell-time based stability conditions for the the system $\Sigma_s$ by employing clock-dependent Lyapunov-Krasovskii functional. To this aim, we define the set 
\begin{equation*}
\mathscr{P}_{\geq T_D} = \bigg\lbrace
\begin{array}{lll}
 \rho : \mathbb{R}_{\geq 0} \rightarrow \mathcal{P} : \rho(t) = \alpha_k \in \mathcal{P}, \\ 
t \in [ t_k, t_{k+1}), t_{k + 1} \geq t_k + T_D, k \in \mathbb{N}_0  
\end{array}
\bigg\rbrace\; ,
\end{equation*}
which contains all the possible parameter trajectories.
\begin{theorem}
\label{thm1}
For given constants $h \geq 0$, $\mu \in [0, 1]$, $\kappa>0$, and  $T_D>0$, if there exist matrix-valued functions $P : [0, \ T_D] \times \mathcal{P} \mapsto \mathbb{S}^n_{\succ 0}$, $Q : [0, \ T_D] \times \mathcal{P} \mapsto \mathbb{S}^n_{\succ 0}$, and $R : [0, \ T_D] \times \mathcal{P} \mapsto \mathbb{S}^n_{\succ 0}$ such that the LMIs
\begin{flalign}
\label{main_result1}
&{
\Gamma(\tau,\rho) := \left[ 
\begin{matrix}
\Gamma_{11}(\tau,\rho) & \Gamma_{12}(\tau,\rho) & hA^T(\rho)R(\tau,\rho) \\ 
* & \Gamma_{22}(\tau,\rho) & h A_d^T(\rho) R(\tau,\rho) \\
* & * & -R(\tau,\rho)  
\end{matrix}
\right] \prec 0 }\\[3mm]
\label{main_result2} 
& \Gamma(T_D^+,\rho) \prec 0  \\[2mm]
\label{main_result4}
& P(T_D,\rho) - P(0,\eta) \succeq 0 \\[2mm]
\label{main_result5}
& Q(T_D,\rho) - Q(0,\eta) \succeq 0  \\[2mm]
\label{main_result6}
& R(T_D,\rho) - R(0,\eta) \succeq 0 \\[1mm]
\label{main_result7}
& \kappa Q(T_D,\rho) - \dot{Q}(0,\eta) \succeq 0\\[1mm]
\label{main_result8}
& \kappa R(T_D,\rho) - \dot{R}(0,\eta) \succeq  0 
\end{flalign}
hold for all  $\tau \in [0, \ T_D]$ and all $\rho, \eta \in \mathcal{P}$,  where 
\begin{equation*}
\begin{array}{lll}
\Gamma_{11}(\tau,\rho) = \emph{Sym}[P(\tau,\rho)A(\rho)] + \dot{P}(\tau,\rho) + Q(\tau,\rho)  
\\ \hspace{5em}-  e^{-\kappa h}R(\tau,\rho)  \\  
\Gamma_{12}(\tau,\rho) = P(\tau,\rho) A_{d}(\rho) + e^{-\kappa h}R(\tau,\rho) \\
\Gamma_{22}(\tau,\rho) = -(1 - \mu) e^{-\kappa h}Q(\tau,\rho) - e^{-\kappa h}R(\tau,\rho)
\end{array}
\end{equation*}
then the system $\Sigma_s$ with $u \equiv 0, w \equiv 0$, and $\rho \in \mathscr{P}_{\geq T_D}$ is uniformly asymptotically stable.    
\end{theorem}

{\em Proof:}\;
Let us define the clock-dependent Lyapunov-Krasovskii functional 
\begin{equation} \nonumber
\label{lyp}
\begin{array}{lll}
V(t,x_t,\rho) &=& x^TP(\tau,\rho)x  
\\[1mm]
&&+ \displaystyle{\int_{t - d(t)}^t} e^{\kappa (s - t)} x^T(s) Q(\tau,\rho) x(s) ds  \\[4mm] 
&&+ h \displaystyle{\int_{-h}^0 \int_{t + \theta}^t} e^{\kappa (s - t)} \dot{x}^T(s) R(\tau,\rho) \dot{x}(s) ds d\theta \; ,
\end{array}
\end{equation}
where $\rho \in \mathscr{P}_{\geq T_D}, \tau = \min \{ t - t_k, T_D \}$, and $t_k$ is the instant where the parameter vector $\rho$ changes its value with a finite jump intensity. 

Taking the derivative of $V$  along the trajectories of  $\Sigma_s$ satisfies
\begin{equation} \nonumber
\label{dervlyp1}
\begin{array}{lll}
\dot{V}(t,x_t,\rho) \preceq 2x^T(t) P(\tau,\rho) x(t) \\[2mm] \hspace{1em} + x^T(t) \dot{P}(\tau,\rho)  
+ x^T(t) Q(\tau,\rho) x(t) \\ [1mm]
\hspace{1em} - (1 - \mu) e^{-\kappa h} x^T(t - d(t)) Q(\tau,\rho) x(t - d(t)) \\[1mm] 
\hspace{1em}+ \displaystyle{\int_{t - d(t)}^t} e^{\kappa (s - t)} x^T(s) [\dot{Q}(\tau,\rho) - \kappa Q(\tau,\rho)] x(s) ds \\[2mm]
\hspace{1em}+ h^2 \dot{x}^T(t)R(\tau,\rho) \dot{x}(t) - h \displaystyle{\int^t_{t - h}} \dot{x}^T(s) e^{-\kappa h} R(\tau,\rho) \dot{x}(s) ds \\[3mm]
\hspace{1em}+ h \displaystyle{\int_{-h}^0 \int_{t + \theta}^t} e^{\kappa (s - t)} \dot{x}^T(s)[ \dot{R}(\tau,\rho) - \kappa R(\tau,\rho)] \dot{x}(s) ds d\theta  \; .
\end{array}
\end{equation}
Since (\ref{main_result7}) and (\ref{main_result8}) hold, it follows that
\begin{equation}
\label{dervLyp}
\begin{array}{lll}
\dot{V}(t,x_t,\rho) \preceq 2 x^T(t) P(\tau,\rho) x(t)  \\[3mm] 
\hspace{1em} + x^T(t) \dot{P}(\tau,\rho) x(t) + x^T(t) Q(\tau,\rho) x(t) \\[3mm] 
\hspace{1em} - (1 - \mu)e^{-\kappa h} x^T(t - d(t)) Q(\tau,\rho) x(t - d(t)) \\[3mm]
\hspace{1em} + h^2 \dot{x}^T(t)R(\tau,\rho) \dot{x}(t)\\[3mm]
\hspace{1em} - h \displaystyle{\int^t_{t - h}} \dot{x}^T(s) e^{-\kappa h} R(\tau,\rho) \dot{x}(s) ds \; .
\end{array}
\end{equation}
Employing Proposition~\ref{prop:Jensen}, we deduce from \rref{dervLyp} that 
\begin{equation}
\label{dervLyp2}
\begin{array}{lll}
\dot{V}(t,x_t,\rho) &\preceq& \xi^T(t) 
\left[
\begin{matrix}
\Gamma_{11}(\tau,\rho) & \Gamma_{12}(\tau,\rho) \\ 
 * &  \Gamma_{22}(\tau,\rho)
\end{matrix}
\right] \xi(t)  \\[5mm] 
&& + h^2 \dot{x}^T(t)R(\tau,\rho) \dot{x}(t)
\; ,
\end{array}
\end{equation}
where 
\[\xi(t)= \left[ 
\begin{matrix}
x^T(t)  & x^T(t - d(t))
\end{matrix}
\right]^T\; .
\]
Taking Schur  compliment of \rref{dervLyp2} yields the LMI (\ref{main_result1}). This condition will ensure that Lyapunov-Krasovskii function is decreasing between two consecutive jumps of the parameter vector $\rho$. Moreover, the change in Lypaunov-Krasovskii functional at the jumping instant $t_k $ of the  parameters' trajectory is given as
\begin{equation*}
\fontsize{9.5}{9.5} \selectfont
\begin{array}{lll}
V(t^-_k,x_t,\rho) - V(t^+_k,x_t,\eta) = x^T(t)[P(T_D,\rho) - P(0,\eta)] x(t) \\[3mm]
+ \displaystyle{\int_{t_k - d(t)}^{t_k}} e^{-\kappa (s - t_k)} x^T(s) [Q(T_D,\rho) - Q(0,\eta)]x(s) ds  \\[5mm] 
+ h \displaystyle{\int_{-h}^{0} \int_{t_k + \theta}^{t_k} e^{- \kappa (s - t_k)} \dot{x}^T(s) [R(T_D,\rho) - R(0,\eta)] \dot{x}(s)} ds d\theta \; .
\end{array}
\end{equation*}
Since (\ref{main_result4}), (\ref{main_result5}), and (\ref{main_result6}) hold, the Lyapunov-Krasovskii functional cannot increase at the time instant $t_k$ as $V(t^-_k,x_t,\rho) - V(t^+_k,x_t,\theta) \succeq 0$. Therefore,  $\Sigma_s$ is uniformly asymptotically stable. This concludes the proof.   \hfill$\square$

\begin{remark}
The Lyapunov-Krasovskii functional is parameter and clock-dependent during the holding time
$t \in [t_k, t_k + T_D]$. For $t > t_k + T_D$, the matrix-valued functions $P(\tau,\rho)$, $Q(\tau,\rho)$, $R(\tau,\rho)$ are chosen to be only parameter dependent such that $P(\tau,\rho) = P(T_D,\rho)$.
\end{remark}

\section{Stabilization with guaranteed $\mathcal{L}_2$-performance by state-feedback}
In this section, we aim at obtaining synthesis conditions for the clock-dependent  gain-scheduled state-feedback controllers of the form
\begin{equation*}
\label{cont}
\Sigma_c:\quad u(t) = \left\lbrace 
\begin{array}{lll}
K(t-t_k,\rho(t_k)) x(t), \; t \in [t_k,t_k+T_D ) \\
K(T_D,\rho(t_k))x(t),   \; t \in [ t_k + T_D, t_{k+1}) \; ,
\end{array}
\right.
\end{equation*}
where $\rho \in \mathscr{P}_{\geq T_D}$, $\tau \in [0, \ T_D]$, $\tau = \min \{ t - t_k, T_D \}$, and $K : [0, \ T_D] \times \mathcal{P} \mapsto \mathbb{R}^{n_u \times n}$ is the clock- and parameter-dependent gain. Our objective is to find  sufficient stabilization conditions for the gain $K(\tau,\rho)$ such that the closed-loop system $(\Sigma_s,\Sigma_c)$ is asymptotically stable in the absence of disturbance $w$ and that the map $w \mapsto z$ has a guaranteed $\mathcal{L}_2$-gain of at most $\gamma$. We have the following the result.
\begin{theorem}
\label{thm2}
For given constants $h \geq 0$, $\mu \in [0, 1]$, ${\kappa>0}$, and $T_D>0$, if there exist matrix-valued functions 
$\tilde{P} : [0, \ T_D] \times \mathcal{P} \mapsto \mathbb{S}^n_{\succ 0}$, $\tilde{Q} : [0, \ T_D] \times \mathcal{P} \mapsto \mathbb{S}^n_{\succ 0}$,  $\tilde{R} : [0, \ T_D] \times \mathcal{P} \mapsto \mathbb{S}^n_{\succ 0}$,  $\tilde{U} : [0, \ T_D] \times \mathcal{P} \mapsto \mathbb{R}^{n_u \times n}$,  and $\tilde{X} :  \mathcal{P} \mapsto \mathbb{R}^{n \times n}$  such that the LMIs (\ref{equ:38a})-(\ref{equ:16}) are feasible: 
\setcounter{equation}{10}
\begin{flalign}
\label{main_result_first_lmi}
&\tilde{\Gamma}(T_D^+,\rho)  \prec 0 \\[1mm]  
& \tilde{P}(T_D,\rho) - \tilde{P}(0,\eta)  \succeq 0 \\[1mm]  
 &\tilde{Q}(T_D,\rho) - \tilde{Q}(0,\eta) \succeq 0  \\[1mm]
 &\tilde{R}(T_D,\rho) - \tilde{R}(0,\eta) \succeq 0  
\end{flalign}
\begin{flalign}
 &\kappa \tilde{Q}(T_D,\rho) - \dot{\tilde{Q}}(0,\eta) \succeq 0  \\[1mm]
 \label{equ:16}
 &\kappa \tilde{R}(T_D,\rho) - \dot{\tilde{R}}(0,\eta)  \succeq 0
\end{flalign}

for all $\tau \in [0, T_D]$ and all $\rho, \eta \in \mathcal{P}$, where 
\begin{equation*}
\begin{array}{lll}  
\tilde{\Gamma}_{12}(\tau,\rho) = \tilde{P}(\tau,\rho)+A(\rho)\tilde{X}(\rho) + B(\rho)\tilde{U}(\tau,\rho)
\\[1mm]
\tilde{\Gamma}_{25}(\tau,\rho) = (C(\rho)\tilde{X}(\rho) + D(\rho)\tilde{U}(\tau,\rho))^T \\[1mm]
\tilde{\Lambda}_{33}(\tau,\rho) = -(1 - \mu) \tilde{Q}(\tau,\rho) e^{-\kappa h} -  e^{-\kappa h} \tilde{R}(\tau,\rho) \\[1mm]
\tilde{\Upsilon}(\tau,\rho) =  \tilde{\dot{P}}(\tau,\rho) + \tilde{Q}(\tau,\rho)  
-  e^{-\kappa h}\tilde{R}(\tau,\rho) - \tilde{P}(\tau,\rho)
\end{array}
\end{equation*}

\begin{figure*}
% ensure that we have normalsize text
\setcounter{equation}{9}
\begin{equation}
\label{equ:38a}
{\fontsize{10}{10} \selectfont
 \tilde{\Gamma}(\tau,\rho) := \left[ 
\begin{matrix}
-\textrm{Sym}[\tilde{X}(\rho)] & \tilde{\Gamma}_{12}(\tau,\rho)  & A_d(\rho)\tilde{X}(\rho) & E(\rho) & 0 & \tilde{X}(\rho) & \tilde{X}(\rho) + h\tilde{R}(\tau,\rho)  \\ 
* & \tilde{\Upsilon}(\tau,\rho) & e^{-\kappa h}\tilde{R}(\tau,\rho) & 0 & \tilde{\Gamma}_{25}(\tau,\rho) & 0  & -\tilde{P}(\tau,\rho)\\ 
* & * & \tilde{\Lambda}_{33}(\tau,\rho) & 0 & \tilde{X}^TC_d^T(\rho) & 0 & 0 \\
* & * & * & -\gamma^2 I & F^T(\rho) & 0 & 0 \\
* & * & * & * & -I & 0  & 0 \\ 
* & * & * & * & * & -\tilde{P}(\tau,\rho) & - h \tilde{R}(\tau,\rho) \\
* & * & * & * & * & * & (-1 - 2h) \tilde{R}(\tau,\rho)
\end{matrix}
\right] \prec 0 }
\end{equation}
\hrulefill
\end{figure*}
then the closed-loop system $(\Sigma_s,\Sigma_c)$ with  $\rho \in \mathscr{P}_{\geq T_D}$ is uniformly asymptotically stable in the absence of
disturbance $w$ and the $\mathcal{L}_2$-gain of the map $w 
\mapsto z$ is at most $\gamma$.  
\end{theorem}  
    
{\em Proof:}\;
From Theorem 1, it follows that
\begin{equation}\nonumber
\begin{array}{lll}
\dot{V}(t,x_t,\rho) &\preceq& \xi^T(t) 
\left[
\begin{matrix}
\Gamma_{11}(\tau,\rho) & \Gamma_{12}(\tau,\rho) \\  * &  \Gamma_{22}(\tau,\rho)
\end{matrix}
\right] \xi(t)  \\[5mm] 
&& + h^2 \dot{x}^T(t)R(\tau,\rho) \dot{x}(t)\; ,
\end{array}
\end{equation}
where $\Gamma_{11}(\tau,\rho)$, $\Gamma_{12}(\tau,\rho)$, and $\Gamma_{22}(\tau,\rho)$ are given in Theorem 1. To ensure the prescribed $\mathcal{L}_2$ performance level  of $\gamma$, we further require 
\setcounter{equation}{16}
\begin{equation}
\label{eq:presprfr}
\dot{V}(t,x_t,\rho) - \gamma^2 w^T(t)w(t) + z^T(t)z(t) \prec 0 \; .
\end{equation}

Substituting $z(t)$ from $\Sigma_s$ in \rref{eq:presprfr} yields  
\begin{equation}
\label{performance}
\zeta^T(t) \Psi(\tau,\rho) \zeta(t) + h^2 \dot{x}^T(t)R(\tau,\rho) \dot{x}(t) \prec 0 \;,
\end{equation}
where 
\begin{equation} \nonumber
\zeta = \left[
\begin{matrix}
x(t) & x(t - d(t)) & w(t)
\end{matrix}
\right]
\end{equation}
and 

${\Psi}(\tau,\rho) = $
\vspace{4pt}
\begin{equation*}\label{equ:38a} 
\begin{array}{lll}
 \left[ 
\begin{matrix}
{\Psi}_{11}(\tau,\rho) &  {\Psi}_{12}(\tau,\rho) & P(\tau,\rho)E(\rho) + C^T(\rho)D(\rho)  \\ 
* & {\Psi}_{22}(\tau,\rho)  & C_d^T(\rho)D(\rho)  \\ 
* &  * & -\gamma^2 I + F^T(\rho)F(\rho) 
\end{matrix}
\right]
\end{array}
\end{equation*}
\\
with 
\vspace{4pt}
\begin{equation} \nonumber
\begin{array}{lll}
\Psi_{11}(\tau,\rho) = \textrm{Sym}[P(\tau,\rho)A(\rho)] + \dot{P}(\tau,\rho) + Q(\tau,\rho)  
\\ \hspace{5em}-  e^{-\kappa h}R(\tau,\rho) + C^T(\rho)C(\rho)  
\\
\Psi_{12}(\tau,\rho) = P(\tau,\rho) A_{d}(\rho) + e^{-\kappa h}R(\tau,\rho)  + C^T(\rho)C_d(\rho) 
\\
\Psi_{22}(\tau,\rho) = -(1 - \mu) e^{-\kappa h} Q(\tau,\rho) - e^{-\kappa h}R(\tau,\rho) \\ 
\hspace{5em} + C_d^T(\rho)C_d(\rho)
\end{array}
\end{equation}
for all  $\rho \in \mathscr{P}_{\geq T_D}$ and all $\tau \in [0, \ T_D]$. 

Applying  Schur complement twice on the LMI \rref{performance}, we obtain

${\Lambda}(\tau,\rho) =$
\begin{equation}
\label{equ:middle}
{\fontsize{9}{9} \selectfont
 \left[ 
\begin{matrix}
{\Lambda}_{11}(\tau,\rho) & {\Lambda}_{12}(\tau,\rho) & {P}(\tau,\rho) E(\rho) & C^T(\rho) &   h A^T(\rho) {R}(\tau,\rho) \\ 
*  & {\Lambda}_{22}(\tau,\rho) & 0 & C_d^T(\rho) &  h A_d^T(\rho){R}(\tau,\rho) \\
* & * & -\gamma^2 I & F^T(\rho) &  h E^T(\rho) {R}(\tau,\rho)  \\
* & * &  * & -I & 0  \\ 
* & * & * & * & - {R}(\tau,\rho)  
\end{matrix}
\right] \prec 0\; , }
\end{equation}
where 
\begin{equation} \nonumber
\begin{array}{lll}
\Lambda_{11}(\tau,\rho) = \textrm{Sym}[P(\tau,\rho)A(\rho)] + \dot{P}(\tau,\rho) + Q(\tau,\rho)  
\\ \hspace{5em}-  e^{-\kappa h}R(\tau,\rho) \\ 
\Lambda_{12}(\tau,\rho) = P(\tau,\rho) A_{d}(\rho) + e^{-\kappa h}R(\tau,\rho)   \\
\Lambda_{22}(\tau,\rho) = -(1 - \mu) e^{-\kappa h} Q(\tau,\rho) - e^{-\kappa h}R(\tau,\rho)\; . 
\end{array}
\end{equation}
The structure of (\ref{equ:middle}) is not adapted to the controller design due to the existence of the multiple product terms $A(\rho)P(\tau,\rho)$ and $A(\rho)R(\tau,\rho)$ that prevent  finding a linearizing change of variable even after congruence transformations. A relaxation approach based on the idea of \cite{briat2010memory} is applied to remove theses multiple product terms as follows. We will first prove that feasibility of (\ref{equ:27a}) guarantees the feasibility of (\ref{equ:middle}). To this aim, we let (\ref{equ:27a}) be called $\hat{\Gamma}(\tau,\rho)$ with $\Upsilon(\tau,\rho) = \dot{P}(\tau,\rho) + Q(\tau,\rho)  -  e^{-\kappa h}R(\tau,\rho) - P(\tau,\rho)$, and decompose it as follows:
\begin{equation}
\hat{\Gamma}(\tau,\rho) = \hat{\Gamma}(\tau,\rho) |_{X = 0} + U^T X V + V^T X^T U\; ,
\end{equation}
where $U = \left[-I_n \; A(\rho) \; A_d(\rho) \; E(\rho)\; \mathcal{O}_n \; I_n  \; I_n  \right] $ and $V = \left[ I_n \; \mathcal{O}_{n \times 6n}
\right]$. Then invoking the projection lemma (\cite{gahinet1994linear}), the feasibility of
$\hat{\Gamma}(\tau,\rho) \prec 0$ implies the feasibility of the LMIs
\begin{subequations}
\begin{equation}
\label{equ:21a}
\mathscr{N}^T_{U} \hat{\Gamma}(\tau,\rho) |_{X = 0} \mathscr{N}_{U} \prec 0 
\end{equation}
\begin{equation} 
\label{equ:21b}
\mathscr{N}^T_{V} \hat{\Gamma}(\tau,\rho) |_{X = 0} \mathscr{N}_{V} \prec 0\; ,
\end{equation}
\end{subequations}
where $\mathscr{N}_{U}$ and $\mathscr{N}_{V}$ are basis of the null space of $U$ and $V$ and given as 
\begin{equation}
\mathscr{N}_U = \left[
\begin{matrix}
A(\rho) & A_d(\rho) & E(\rho) & 0 & I & I  \\ 
I & 0 & 0 & 0 & 0 & 0 \\ 
0 & I & 0 & 0 & 0 & 0   \\ 
0 & 0 & I & 0 & 0 & 0  \\ 
0 & 0 & 0 & I & 0 & 0  \\
0 & 0 & 0 & 0 & I & 0  \\
0 & 0 & 0 & 0 & 0 & I   
\end{matrix}
\right], \ 
\mathscr{N}_V = \left[
\begin{matrix}
0 \\ 0 \\ 0 \\ 0 \\ 0 \\ 0  \\ I
\end{matrix} 
\right]\; .
\end{equation}

Subsequently, the projection lemma yields two inequalities, where  the first inequality \rref{equ:21a} yields (\ref{equ:middle}) and \rref{equ:21b} yields $-R(\tau,\rho) \prec 0$ for all $\rho \in \mathscr{P}_{\geq T_D}$ and all $\tau \in [0, T_D]$. Note that this inequality is a relaxed form of the right bottom $1 \times 1$ block of the inequality (\ref{equ:middle}) and is always satisfied. Hence, the feasibility of (\ref{equ:27a}) implies
the feasibility of (\ref{equ:middle}).

\begin{figure*}
% ensure that we have normalsize text
\begin{equation}
\label{equ:27a}
{\fontsize{9}{9} \selectfont
\hat{\Gamma}(\tau,\rho) :=  \left[ 
\begin{matrix}
-\textrm{Sym}[X(\rho)] & P(\tau,\rho)+X^T(\rho)A(\rho)  & X^T(\rho)A_d(\rho) & X^T(\rho)E(\rho) & 0 & X^T(\rho) & X^T(\rho) + hR(\tau,\rho)  \\ 
* & \Upsilon(\tau,\rho) & e^{-\kappa h}R(\tau,\rho) & 0 & C^T(\rho) & 0  & -P(\tau,\rho)\\ 
* & * & \Lambda_{22}(\tau,\rho) & 0 & C_d^T(\rho) & 0 & 0 \\
* & * & * & -\gamma^2 I & F^T(\rho) & 0 & 0 \\
* & * & * & * & -I & 0  & 0 \\ 
* & * & * & * & * & -P(\tau,\rho) & - h R(\tau,\rho) \\
* & * & * & * & * & * & (-1 - 2h)R(\tau,\rho)
\end{matrix}
\right] \prec 0 }
\end{equation}
% Restore the current equation number.
%\setcounter{equation}{\value{MYtempeqncnt}}
% IEEE uses as a separator
% The spacer can be tweaked to stop underfull vboxes.
\hrulefill
\end{figure*}

Finally, for the controller synthesis, substituting the closed-loop matrices 
\begin{equation} \nonumber
\begin{array}{rcl}
A(\rho) \leftarrow  A_{cl}(\tau,\rho) & :=& A(\rho) + B(\rho)K(\tau,\rho)
\\
 C(\rho) \leftarrow  C_{cl}(\rho) &:=& C(\rho) + D(\rho)K(\tau,\rho)
\end{array}
\end{equation}
into the inequality (\ref{equ:27a}), and then performing a congruence transformation with respect to matrix {$diag(X^{-1}(\rho)$, $X^{-1}(\rho)$, $X^{-1}(\rho)$,$I$,$I$, $X^{-1}(\rho)$,$X^{-1}(\rho))$} along with the linearizing change of variables
\begin{equation}
\begin{array}{rcl}
\tilde{X}(\rho) &:=& X^{-1}(\rho)
\\
\tilde{P}(\rho) &:=& \tilde{X}^T(\rho) P(\tau,\rho) \tilde{X}(\rho)
\\
\tilde{R}(\tau,\rho) &:=& \tilde{X}^T(\rho) R(\tau,\rho) \tilde{X}(\rho)
\\
\tilde{\Lambda}_{22}(\tau,\rho) &:=& \tilde{X}^T(\rho) \Lambda_{22}(\tau,\rho) \tilde{X}(\rho)
\\
\tilde{\Upsilon}(\tau,\rho)& :=& \tilde{X}^T(\rho) \Upsilon(\tau,\rho) \tilde{X}(\rho)
\\
\tilde{U}(\tau,\rho) &:=& K(\tau,\rho)\tilde{X}^{-1}(\rho)
\\
\tilde{\dot{P}}(\tau,\rho) &:=& \tilde{X}^T(\rho) \dot{P}(\tau,\rho) \tilde{X}(\rho)
\end{array}
\end{equation}
yield the LMI \rref{equ:38a}. This concludes the proof.  \hfill$\square$
\section{Illustrations}
 We now provide three examples. The purpose of first example is to illustrate Theorem~\ref{thm1} whereas the second one illustrates  Theorem~\ref{thm2}. The third example demonstrates the application of our results to the consensus problem of multi-agent systems.
\subsection{Example 1: Illustration of Theorem~\ref{thm1}}
 Let us consider the following LPV system with time delay considered in \cite{pang2015stability}:
\begin{equation}
\label{eq:example1}
\begin{array}{lll}
\dot{x}(t) = \left[
\begin{matrix}
0 & 1 \\ -2-\rho & -1  
\end{matrix}
\right]x(t) + \left[
\begin{matrix}
-1 & 0 \\ -1-\rho & -1
\end{matrix}
\right]x(t - d(t))\;,
\end{array}
\end{equation} 
where $d(t) < 0.5$ and $\dot{d}(t) \leq \mu < 0.5$. We solve the LMIs in Theorem \ref{thm1} via gridding approach with fifty points in YALMIP, \cite{lofberg2004yalmip}. Since the  LMIs in Theorem \ref{thm1} yield intractable infinite-dimensional semi-definite programs, we relax them  by using parameter-dependent polynomials of order 1.  Choosing  $T_D = 1 \times 10^{-4}$ and $\kappa =   0.005$, and applying Theorem~1, one can corroborate that the system \rref{eq:example1} is stable for $0 < \rho(t) \leq 0.76$.
\subsection{Example 2: Illustration of Theorem~\ref{thm2}} 
We now consider the system $\Sigma_s$ with 
\begin{equation}
\label{sys-ex2}
\begin{array}{lll}
\fontsize{9}{9} \selectfont
A(\rho) = \left[
\begin{matrix}
2 - \rho & -0.5 - 0.5 \rho \\ -1 & -2+0.1\rho
\end{matrix}
\right], \;
A_d(\rho) = \left[
\begin{matrix}
-1 & 0 \\ 0.05-0.45\rho & -1
\end{matrix}
\right], \\[4mm] 
\fontsize{9}{9} \selectfont
B(\rho) = \left[ 
\begin{matrix}
1 \\ 0 
\end{matrix}
\right], \;
E(\rho) = \left[
\begin{matrix}
0.01 \\ 0.01
\end{matrix}
\right], \;
C(\rho) = C_d(\rho) = \left[
\begin{matrix}
0 & 1
\end{matrix}
\right], \\[3mm]
D(\rho) = F(\rho) = 0, \; \mathcal{P} = [0,\; \bar{\rho}]  \;.
\end{array}
\end{equation}

\begin{figure}
\begin{center}\vspace{.5em}
\includegraphics[width =\linewidth]{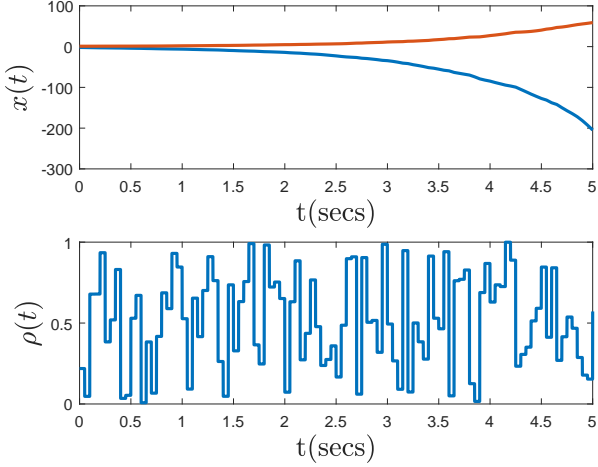}\vspace{-.5em}
\caption{State trajectories of the open-loop system (top) and a
typical parameter trajectory (bottom)}
\label{fig:Fig1}
\end{center}\vspace{-.5em}
\end{figure}

\begin{figure}
\begin{center}\vspace{.5em}
\includegraphics[width =\linewidth]{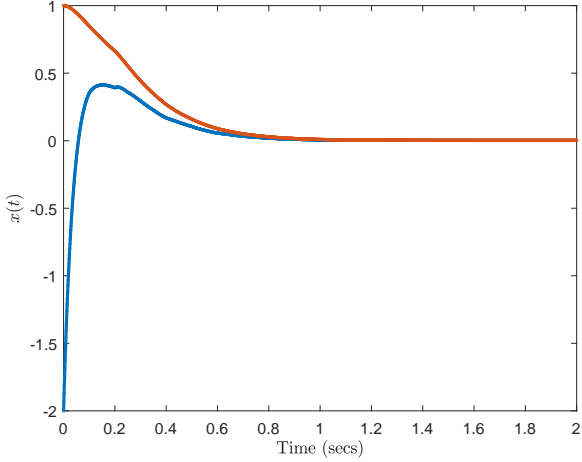}\vspace{-.5em}
\caption{State trajectories of the closed-loop system under a step disturbance}
\label{fig:Fig2}
\end{center}\vspace{-.5em}
\end{figure}

Choosing $\bar{\rho} = 1$, $h = 0.2$, $\mu = 0.9$, $\kappa =  1 \times 10^{-8}$, and $T_d = 0.01$, and solving the LMIs in Theorem~2 yields the following controller gain
  \begin{equation} \nonumber
  \label{kontex2}
  K(\tau,\rho) = \frac{1}{den(\tau,\rho)} \left[
  \begin{matrix}
  K_1(\tau,\rho) & K_2(\tau,\rho)
  \end{matrix}
  \right]
  \end{equation}
  where 
  \begin{equation}\nonumber
  \begin{array}{lll}
  K_1(\tau,\rho) =  2031.1\rho + 2364.7\tau + 6.8217\rho\tau  - 46431.0 \\
  K_2(\tau,\rho) =10.565\rho\tau - 2181.9\tau - 2588.6\rho + 26748.0 \\
  den(\rho) = 47.187\rho + 1639.5 \; .
  \end{array}
  \end{equation}
We simulate both the open-loop system and the closed-loop system under a  unit-step disturbance. It can be observed in Fig.~\ref{fig:Fig1} (top) that the open-loop system is unstable. At the bottom of the same figure, a random parameter trajectory is shown, where the time between two successive jumps is taken equal to $T_D=0.05$ and the next value for the parameter is simply drawn from $\mathcal{U}(0,2)$. The evolution of the closed-loop state trajectories is shown in Fig. \ref{fig:Fig2} with an initial condition $[-2, 1]$. Since our simulation depicts stability of the closed-loop system, it helps illustrate our general theory, in the special case of the system \rref{sys-ex2}.

\subsection{Example 3: Consensus problem of multi-agent systems}
In this example, we illustrate the application of our results to consensus problem of a multi-agent nonholonomic system subject to a switching topology.   

Consider the following multi-agent system with   six agents ($N = 6$) considered in \cite{gonzalez2014lpv}:

\begin{equation}
\fontsize{9.5}{9.5} \selectfont
\label{sys_def_ex3}
\begin{array}{lll}
\dot{\bar{x}}(t) &=& (I_N \otimes \mathcal{A})\bar{x}(t) + (\mathcal{L}(t) \otimes \mathcal{A}_d)\bar{x}(t - d(t)) \\ 
&&+ (I_N \otimes \mathcal{B} ) \bar{u}(t) + (I_N \otimes \mathcal{E}) \bar{w}(t) \\ 
\bar{z}(t) &=& (I_N \otimes \mathcal{C}) \bar{x}(t) + (\mathcal{L}(t) \otimes \mathcal{C}_d) \bar{x}(t - d(t)) \\
&&+ (I_N \otimes \mathcal{D})\bar{u}(t) + (I_N \otimes \mathcal{F}) \bar{w}(t) 
\end{array}
\end{equation} 
where $\bar{x}(\theta) = \phi(\theta)$ for all $\theta \in \left[-h, 0  \right]$, $\bar{x}\in \mathbb{R}^{Nn}$, $\bar{u} \in \mathbb{R}^{Nq}$,  $\bar{w} \in \mathbb{R}^{Nm}$, $\bar{z} \in \mathbb{R}^{Nr}$, and the system matrices are given by
\begin{equation}
\begin{array}{lll}
\mathcal{A} = 
\left[
\begin{matrix}
0 & 1 \\ -1 & 0 
\end{matrix}
\right], \
\mathcal{A}_d = 
\left[
\begin{matrix}
0 & 1 \\ -1 & 0 
\end{matrix}
\right], \
\mathcal{B} = 
\left[
\begin{matrix}
1 & 0 \\ 0 & 0.6
\end{matrix}
\right], \\
\mathcal{E} =  0.05 \times \mathcal{B}, \
\mathcal{C} = \mathcal{C}_d = I_2, \ 
\mathcal{D} = \mathcal{O}_2, \  \mathcal{F} = 0.1 I_2 \; .
\end{array}
\end{equation} 
The upper bounds on delay and its derivative are chosen to be $h = 0.2$ and $\mu = 0.9$, respectively. For the system  \rref{sys_def_ex3}, we consider a switching topology represented by the following  time-varying Laplacian matrix
\begin{equation}
\begin{array}{rcl}
\mathcal{L}(t) = \sigma(t)\mathcal{L}_1 + (1 - \sigma(t))\mathcal{L}_2\; ,
\end{array}
\end{equation}
where $\sigma(t)$ is a piecewise constant switching signal that takes value in $[0,1]$ with $T_D = 0.1$, and  $\mathcal{L}_1, \mathcal{L}_2$ is a pair of symmetric commutable Laplacian matrices given by 
\begin{equation}
\fontsize{9}{9} \selectfont
\begin{array}{rcl}
\mathcal{L}_1 &=& 
\left[
\begin{matrix}
1 &-0.5 & 0& 0& 0 &-0.5 \\ 
    -0.5 & 1& -0.5& 0 &0& 0\\
    0 &-0.5& 1 &-0.5& 0 &0 \\
    0 &0 &-0.5 & 1 &-0.5& 0  \\ 
    0 &0 &0 &-0.5& 1 &-0.5 \\
    -0.5& 0& 0 &0 &-0.5 &1  \\
\end{matrix}
\right], 
\end{array}
\end{equation}
\begin{equation}
\fontsize{9}{9} \selectfont
\begin{array}{rcl}
\mathcal{L}_2 &=&
\left[
\begin{matrix}
1 & -0.25 &-0.25& 0& -0.25 & -0.25  \\
    -0.25& 1 &-0.25 &-0.25 & 0 & -0.25 \\ 
    -0.25 &-0.25& 1 &-0.25 & -0.25 & 0   \\
    0 &-0.25 &-0.25& 1 & -0.25 & -0.25 \\
    -0.25& 0& -0.25 &-0.25 & 1  & -0.25   \\   
    -0.25 &-0.25& 0 &-0.25 &-0.25& 1 \\
\end{matrix}
\right].
\end{array}
\end{equation}
For the  matrices $\mathcal{L}_1$ and $\mathcal{L}_2$, the maximum and minimum eigenvalues are  $\min\{\underline{\lambda}_1,\ \underline{\lambda}_2 \}= 0$, $\max\{\bar{\lambda}_1,\ \bar{\lambda
}_2 \} = 2$, respectively.  By defining a new piecewise constant parameter $\rho \in \mathscr{P}_{\geq T_D}$ with  $ \mathcal{P} = [0, \ \bar{\rho}]$ where $\bar{\rho} = 2$ and using an approach similar to Corollary 2 of \cite{zakwan2020onoutput}, the distributed system \rref{sys_def_ex3} can be modeled as following LPV system with piecewise constant parameter:
\begin{equation}
\fontsize{9.5}{9.5} \selectfont
\label{sys_def3}
\begin{array}{lll}
\dot{x}(t) &=& A(\rho)x(t) + A_d(\rho)x(t - d(t)) + B(\rho) u(t) + E(\rho) w(t) \\ 
z(t) &=& C(\rho)x(t) + C_d(\rho)x(t - d(t)) + D(\rho)u(t) + F(\rho)w(t) \\ 
x(\theta) &=& \phi(\theta), \ \forall \theta \in \left[-h, 0  \right]
\end{array}
\end{equation}
where $ {A}(\rho) = \mathcal{A}$,  ${A}_d(\rho) = \rho \mathcal{A}_d$,  $B(\rho) = \mathcal{B}$,  $E(\rho) = \mathcal{E}$,  ${C}(\rho) = C$,  $C_d(\rho) = \rho\mathcal{C}_d$, $D(\rho) = \mathcal{D}$, and $F(\rho) = \mathcal{F}$. 

Our goal is to design a clock-dependent distributed controller $\mathcal{K}(\tau,t) = I_N \otimes \mathcal{K}_a(\tau) + \mathcal{L}(t) \otimes \mathcal{K}_b(\tau)$. Using an approach similar to Corollary 2 of \cite{zakwan2020onoutput}, the distributed controller $\mathcal{K}(t)$ can be modelled as the clock-dependent gain-scheduled controller $\Sigma_c$ with gain ${K}(\tau,\rho)= K_a(\tau) + \rho K_b(\tau)$. We Choose $\kappa = 0.01$ and then solve the LMIs in Theorem~2 using gridding approach with fifty points in YALMIP, \cite{lofberg2004yalmip}. Note that to inherit same parametrization for controller as of the plant, we used $\tilde{U}(\tau, \rho) = \tau\tilde{U}_a + \rho \tilde{U}_b$.  Once LMIs are feasible, we compute the matrix-valued function $\tilde{U}(\tau,\rho)$ which results in the following controller gains
\begin{equation}
\begin{array}{lll}
K_a(\tau) = \left[ 
\begin{matrix}
 -11.102\tau & -11.573\tau \\
        17.813\tau & -14.79\tau
\end{matrix}
\right] \\ 
K_b(\tau) = \left[ 
\begin{matrix}
-0.5078 &-0.63061  \\
       0.87181 & -0.78384 
\end{matrix}
\right] \; .
\end{array}
\end{equation} 
The distributed controller gain $\mathcal{K}(t)$ can be constructed from $\mathcal{K}_a(\tau)=K_a(\tau)$ and $\mathcal{K}_b(\tau)=K_b(\tau)$. 

\begin{figure}
\includegraphics[width = \linewidth]{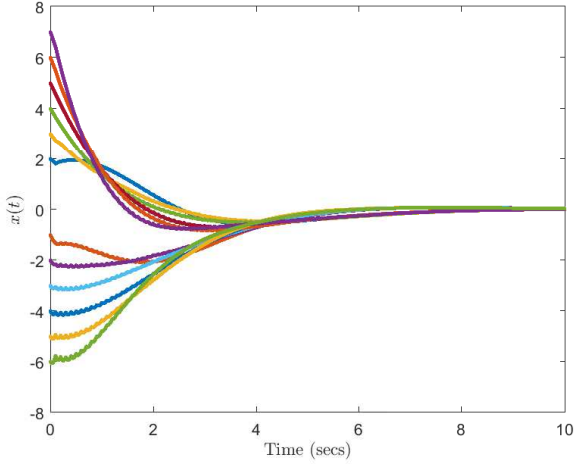}
\caption{State trajectories  of the closed-loop system under a step disturbance}
\label{fig:states}
\end{figure}

\begin{figure}
\includegraphics[width = \linewidth]{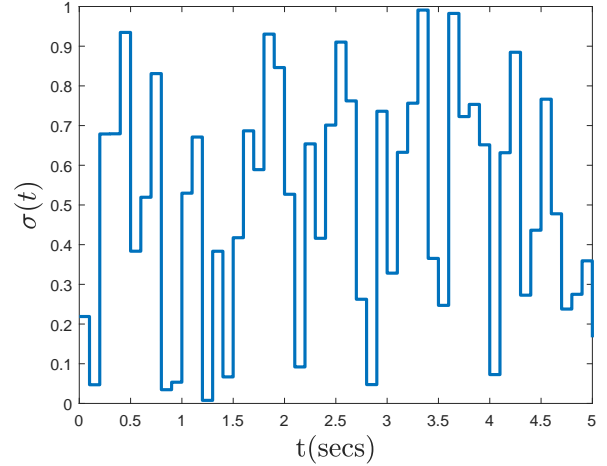}
\caption{A typical realization of the piecewise constant switching signal $\sigma(t)$}
\label{fig:sigma}
\end{figure}

We simulate the closed-loop system with a time-varying delay $\tau(t) = 0.09 sin(0.9t) + 0.1$ under a unit-step disturbance. Fig.~\ref{fig:states} shows the evolution of the state trajectories reaching a consensus subject to a typical switching signal shown in Fig. \ref{fig:sigma}. For the random parameter trajectory shown in Fig.~\ref{fig:sigma}, the time between two successive jumps is taken equal to $T_D=0.1$ and the next value for the parameter is simply drawn from $\mathcal{U}(0,1)$.  The consensus of multiagent system depicted in Fig. \ref{fig:states} under a switching topology and time-varying delay reflects the efficacy of the approach. 
 
\section{Concluding Remarks}
Dwell-time based stability conditions and synthesis conditions for gain-scheduled $\mathcal{L}_2$ state-feedback controllers are derived for a class of LPV systems with piecewise parameters under time-varying delay.  One of the main advantages of our approach is reduced conservatism with improved performance  as compared to more generalized methods such as quadratic stability.     

Several extensions of this work are possible; for instance,  dynamic $\mathcal{L}_2$ output feedback control,  improving the bound on the rate of change of time delay, considering stochastic time-delays and stochastic piecewise constant parameter trajectories.     

\begin{ack}
The authors would like to thank Dr. Corentin Briat for his useful suggestions to improve the quality of this paper.
\end{ack}

\bibliographystyle{automatica}
\bibliography{mybibfile}

\end{document}